\documentclass[a4paper,11pt]{article}

\usepackage{jheppub} 

\usepackage[T1]{fontenc} 
\usepackage{bm}

\newcommand{\mev}{\text{ MeV}}

\title{\boldmath Isosinglet-isotriplet mixing and the $X(3872)$ lineshape}

\author[a,b,1]{D. Germani,\note{Corresponding author.}}
\author[c]{B. Grinstein}

\affiliation[a]{INFN Sezione di Roma, Piazzale Aldo Moro 2, I-00185 Rome, Italy}
\affiliation[b]{Sapienza University of Rome, Piazzale Aldo Moro 2, I-00185, Italy}
\affiliation[c]{University of California, San Diego, 9500 Gilman Drive, La Jolla, CA 92093, USA}

\emailAdd{davide.germani@uniroma1.it}
\emailAdd{bgrinstein@ucsd.edu}

\abstract{We investigate the lineshapes of the $X(3872)$ in $B^+$ decays production within a framework that incorporates two underlying QCD configurations: a compact isosinglet state $X_S$ and the neutral component of a molecular isotriplet $X_T^0$. The physical signal is interpreted as arising from the mixing of these states, induced by strong isospin breaking. The decay amplitude is constructed in a factorized form, separating short-distance production, non-relativistic propagation and final-state interactions. This setup allows for a unified description of both $DD^*$ and $J/\psi\,+$ pions final states. We show that the interplay between the two components and their mixing can qualitatively reproduce several nontrivial experimental features. In particular, interference effects can enhance the charged $DD^*$ channel relative to the neutral one despite phase-space suppression, and generate distinctive structures in the $J/\psi \pi^+\pi^-$ and $J/\psi \pi^+\pi^-\pi^0$ lineshapes, including the possibility of strong distortions near threshold.}

\begin{document} 
\maketitle
\flushbottom

\section{Introduction}
\label{sec:intro}
Since its discovery, the $X(3872)$ ($J^{PC}=1^{++}$), reported in the PDG as $\chi_{c1}(3872)$ \cite{ParticleDataGroup:2024cfk}, has attracted significant attention as the prime candidate for a tetraquark state, exhibiting several peculiar properties. Its mass lies extremely close to the $D^0\bar{D}^{*0}$ threshold~\cite{ParticleDataGroup:2024cfk}; indeed, within current experimental resolution, it is compatible with a vanishing binding energy. Searches for charged partners --- for instance in the $J/\psi \pi^\pm\pi^0$ channel --- have so far yielded negative results \cite{BaBar:2004cah}, leading to the conventional assignment of the $X(3872)$ as an isosinglet state. However, recent results may challenge this assignment. Experimental observations~\cite{Belle:2005lfc,BESIII:2019esk,BaBar:2010wfc}
\begin{equation}
    {\cal I}= \frac{{\text{Br}}(X\to J/\psi\,\pi^+\pi^-\pi^0)}{{\text{Br}}(X\to J/\psi\,\pi^+\pi^-)}=\begin{cases}
        1.0\pm0.4\,\text{(stat)}\pm0.3\,\text{(syst)} & \text{[Belle]}\\
        0.8\pm0.3 & \text{[BABAR]}\\
         1.43^{+0.28}_{-0.23} & \text{[BESIII]}
    \end{cases}\,,
    \label{eq:br_ratioI}
\end{equation}
when interpreted as dominated by $\omega$ and $\rho$ resonances, suggest strong isospin violation. Furthermore, a recent study by the LHCb collaboration~\cite{LHCb:2022jez} measured the ratio of the isospin-violating to isospin-conserving $X(3872)$ couplings as
\begin{equation}
    \mathcal{G}=\frac{g_{X(3872)\to J/\psi\rho^0}}{g_{X(3872)\to J/\psi \omega}} = 0.29 \pm 0.04\,,
    \label{eq:grho_gomega}
\end{equation}
and compared it with the case of the $\psi(2S)$, where
\begin{equation}
    \frac{g_{\psi(2S)\to J/\psi\pi^0}}{g_{\psi(2S)\to J/\psi \eta}} = 0.045 \pm 0.001\,.
    \label{eq:psi-I-conserve}
\end{equation}

In a previous work~\cite{Carducci:2025jed}, we analyzed this strong isospin violation, demonstrating that it can be accommodated by assuming the $X(3872)$ peak originates from two distinct QCD states: a compact isosinglet and the neutral partner of a molecular isotriplet. In Ref.~\cite{Carducci:2025jed}, we showed that the experimental data are compatible with this hypothesis, provided the binding energy $B$ of the bound state lies in the range $B\in[0.1,1]\mev$. The possibility of an isotriplet partner was also investigated in Ref.~\cite{Ji:2025hjw}, where both the isosinglet and isotriplet configurations were treated as molecular states. 

We now turn our attention to an additional observable that may be interpreted as a signature of two resonances underlying the $X(3872)$ signal. Ref.~\cite{LHCb:2024vfz} studied the $B^+\to K^+D^{\pm}D^{*\mp}$ decay, finding that a $1^{++}$ contribution is required to describe the spectrum near the $D^{\pm}D^{*\mp}$ threshold. This contribution, associated with the $X(3872)$, yields the measurement $\text{Br}(B^+\to K^+ X)\text{Br}(X\to D^{+}D^{*-})= (1.48^{+0.41}_{-0.35})\times 10^{-4}$. Comparing this with the neutral channel result $\text{Br}(B^+\to K^+ X)\text{Br}(X\to D^{0}\bar{D}^{*0},\bar{D}^{0}D^{*0})= (0.80\pm0.23)\times 10^{-4}$ \cite{Belle:2008fma, ParticleDataGroup:2024cfk}, we observe an enhancement in the charged mode. Given that the charged channel is phase-space suppressed relative to the neutral one, this counter-intuitive behavior warrants a more detailed investigation.

In this work, we extend the analysis of Ref.~\cite{Carducci:2025jed} by further investigating the consequences of two possible components within the $X(3872)$ structure, denoted as $X_S$ (isosinglet) and $X_T^0$ (isotriplet), which mix due to strong isospin breaking. We incorporate the production mechanism via $B^+$ meson decay, extend the previous discussion on $J/\psi \pi^+\pi^-$ strong isospin breaking and include both neutral and charged $DD^*$ decays, specifically addressing the tension observed in \cite{LHCb:2024vfz}.

The paper is organized as follows: in Section \ref{sec:th_mod}, we describe the theoretical model and its underlying hypotheses, detailing the components of the scattering amplitude for $B^+\to K^+(X\to \text{final states})$. In Sec.~\ref{sec:decayDD} of this study, the $\text{final states}$ (hereafter abbreviated as $\text{f.s.}$) are restricted to $(D^{0}\bar{D}^{*0},\,D^{+}D^{*-})$,\footnote{Throughout the paper  adding charge conjugate to final states  is implied but not explicitly written, {\it e.g.}, $\text{f.s.}=D^{0}\bar{D}^{*0}$ stands for $\text{f.s.}=(D^{0}\bar{D}^{*0}+\bar{D}^{0} D^{*0})/\sqrt{2}$.} and in Sec.~\ref{sec:decayJpsi} to $(J/\psi\pi^+\pi^-,\,J/\psi\pi^+\pi^-\pi^0)$; in Section \ref{sec:decayJpsi}, we present also the differential lineshape, which can be employed for a re-analysis of $B^+\to K^+ J/\psi \pi^+\pi^-$ data \cite{LHCb:2020xds}, as well as for the simultaneous analysis of isospin-preserving and isospin-violating decays, following the approach in \cite{LHCb:2022jez}.

\section{Theoretical model}
\label{sec:th_mod}
In this section, we introduce the model used to study the lineshapes of the $B^+\to K^+ (X\to\text{f.s.})$ decays, which incorporates both a compact isosinglet state $X_S$ and the neutral component of a molecular isotriplet $X_T^0$. This framework extends the discussion presented in Ref.~\cite{Carducci:2025jed} regarding the isospin violation observed in $X(3872)\to J/\psi(\rho,\omega)$ decays~\cite{Carducci:2025jed,LHCb:2022jez}. The core assumption is that the $B^+$ meson decay amplitude can be factorized into three distinct components:
\begin{equation}
    \mathcal{A}_a(B^+\to K^+(X\to\text{f.s.))}=\sum_{i,\,j} \mathcal{D}^{\text{decay}}_{ai}\,\Pi_{ij}(E)\,\mathcal{C}_j^{\text{prod}}
    \label{eq:A_tot}
\end{equation}
where:
\begin{itemize}
    \item $\mathcal{C}^{\text{prod}}_j$, where $j=1,2$ denote the isosinglet and isotriplet respectively, represents the production amplitude of $X_j$ in the $B^+\to K^+ X_j$ process. It should be noted that the $X_j$ produced at the vertex correspond to the states of definite isospin, rather than the physical states which appear as poles of the full amplitude. We parameterize this term using two short-distance form factors
    \begin{equation}
        \mathcal{C}^{\text{prod}}(s)=\begin{pmatrix}
            c_S(s)\\
            c_T(s)
        \end{pmatrix}\,,
    \end{equation}
    where $s$ is the invariant mass of the $K^+X_j$ pair. The energy dependence is negligible since the $B^+$ is essentially produced on-shell, allowing us to fix $s\simeq m_{B^+}^2$~\cite{LHCb:2020xds,LHCb:2024vfz}.
    {Data on prompt $X$ production~\cite{CMS_pp_ptproduction,LHCb_High_mult_pp_collision,CMS_pbpb_ptproduction} disfavor a purely molecular interpretation~\cite{Esposito:2015fsa,Bignamini:2009sk}, strongly suggesting that the short-distance production mechanism must proceed primarily through a compact state. In~\cite{Brambilla:2026ujo}, it is argued that the prompt production of the $X(3872)$ has to originate from a compact color-octet $c\bar{c}$ core. Within our framework, this constraint implies that the initial production is dominated by the compact isosinglet component, allowing us to safely set $c_T \simeq 0$.}\footnote{This also implies that the production of the charged partners $X_T^\pm$ is suppressed, providing a rationale for their lack of experimental observation.}
    \item $\Pi_{ij}(E)$ is the matrix containing the non-relativistic propagators and the mixing between $X_S$ and $X_T^0$. The energy $E$ flowing through the propagators is measured relative to the $D^0\bar{D}^{*0}$ threshold, such that $m_{\text{f.s.}}=E+m_{D^0}+m_{D^{*0}}$. A detailed discussion of $\Pi(E)$ is provided in Sec.~\ref{sec:Pi(E)}.
    \item $\mathcal{D}_{ai}^{\text{decay}}$ is the matrix parameterizing the decay and any potential final-state interactions. 
    The specific form of $\mathcal{D}^{\text{decay}}$ depends on whether it is the $DD^*$ ($a=1,2$ means $(D^0\bar{D}^{*0},D^{+}D^{*-})$) or $J/\psi+\text{pions}$ ($a=1,2$ means $(J/\psi\,\pi^+\pi^-\pi^0,J/\psi\,\pi^+\pi^-)$) final state that is being considered. It is important to note, and we emphasize that the previous two factors, $\mathcal{C}$ and $\Pi$,  are common to both  decays so that this third factor carries all the decay process dependence; this is detailed in Sec.~\ref{sec:decay}. 
    \end{itemize}

It is worth commenting briefly on the factorization proposed in \eqref{eq:A_tot}. The robustness of this factorization has been previously discussed in the context of the operator product expansion \cite{Braaten:2006sy,Artoisenet:2010va,Braaten:2005jj} and applied to $X(3872)$ decays. Additionally, this result can be recovered using the narrow width approximation. This appears to be a very decent approximation; since the $X(3872)$ is narrow \cite{ParticleDataGroup:2024cfk}, we expect the widths involved in this case to be no larger.

\subsection{\texorpdfstring{Propagator matrix: $\Pi(E)$}{Propagator matrix: Pi(E)}}
\label{sec:Pi(E)}
To account for the mixing between the isosinglet and isotriplet states, we write the inverse propagator matrix $\Pi^{-1}(E)$ as~\cite{Carducci:2025jed,Carducci:2026fbz}
\begin{multline}
    \Pi^{-1}(E)=\\\begin{pmatrix}
        E-m_S-\dfrac{m}{2\pi}g_S^2\left(\kappa_0(E)+\kappa_\pm(E)\right)+i\dfrac{\Gamma_S}{2} & g_{\text{mix}}\\
         g_{\text{mix}} & -E+m_T-\dfrac{m}{2\pi}g_T^2\left(\kappa_0(E)+\kappa_\pm(E)\right)+i\dfrac{\Gamma_T}{2}
    \end{pmatrix}\,,
\end{multline}
where $m$ is the reduced mass of the $D^0\bar{D}^{*0}$ pair\footnote{We assume the reduced mass of the neutral channel to be equal to that of the charged one, given that their actual difference is negligibly small: $(m_\pm - m)/m \simeq 0.2\%$.}, and $E$ represents the non-relativistic energy measured relative to the $D^0\bar{D}^{*0}$ threshold. Denoting the final-state invariant mass as $m_{\text{f.s.}}$, the energy is defined as $E = m_{\text{f.s.}} - m_{D^0} - m_{D^{*0}}$. Similarly, the bare mass parameters $m_{S,T}$ are also defined relative to this same reference threshold. Moreover
\begin{equation}
    \kappa_0(E)=\sqrt{-2mE}\,,\quad \kappa_\pm(E)=\sqrt{-2m(E-\Delta)}\,,
\end{equation}
with $\Delta=(m_{D^{+*}}+m_{D^-})-(m_{D^{0*}}+m_{D^0})\simeq8\,\text{MeV}$. The molecular nature of the $X^0_T$ state is implemented by adopting a kinetic term with an inverted sign\footnote{Since the fields are non-relativistic, they have mass dimension $[X_i]=M^{3/2}$. Furthermore, $X_i$ ($X^{\dagger}_i$) contains only particle annihilation (creation) operators.}~\cite{Esposito:2025hlp,Kaplan:1999qa,Kaplan:1996nv,Carducci:2026fbz}
\begin{equation}
\mathcal{L}_{\text{kin}}\supset X_S^\dagger\left(i\partial_t+\frac{\nabla^2}{2m_{X_S}}-m_S\right)X_S-X_T^{0^\dagger}\left(i\partial_t+\frac{\nabla^2}{2m_{X_T}}-m_T\right)X_T^{0}\,.
\end{equation}
The couplings $g_i$ are defined by the isospin symmetric interaction Lagrangian~\cite{Esposito:2025hlp}
\begin{equation}
    \mathcal{L}_{\text{int}}^S=g_S X_S^\dagger\left(\frac{D^0\bar{D}^{*0}+\bar{D}^0D^{*0}}{\sqrt{2}}-\frac{D^+D^{*-}+D^-D^{*+}}{\sqrt{2}}\right)+\text{h.c.}\,,
    \label{eq:lag_S}
\end{equation}
\begin{equation}
    \mathcal{L}_{\text{int}}^T=g_T X_T^{0\dagger}\left(\frac{D^0\bar{D}^{*0}+\bar{D}^0D^{*0}}{\sqrt{2}}+\frac{D^+D^{*-}+D^-D^{*+}}{\sqrt{2}}\right)+\text{h.c.}\,.
    \label{eq:lag_T}
\end{equation}
In our previous work \cite{Carducci:2025jed}, isospin violation was assumed to be driven exclusively by the mass difference between the charged $D^{\pm}(D^{*\pm})$ and neutral $D^0(D^{*0})$ mesons, which allowed us to constrain the value of $g_{\text{mix}}$. However, this is admittedly a simplification, as it neglects other potential sources of isospin breaking, such as loop contributions involving $\rho$-$\omega$ mixing and electromagnetic effects. Consequently, in the present analysis we relax this assumption and treat $g_{\text{mix}}$ as a free parameter. Once its value is extracted from experimental data, comparing it against theoretical predictions will allow us to estimate the actual contribution of $DD^*$ loops to the overall isospin breaking.

Furthermore, within the shallow bound-state approximation, the coupling $g_T$ is related to the binding energy $B$ of the molecule via~\cite{landau,DWBA,Carducci:2026fbz}
\begin{equation}
    g_T^2(B)=\frac{2\pi}{m}\sqrt{\frac{2B}{m}}\,.
\end{equation}
The binding energy appearing in this formula refers to the isotriplet eigenstate, which we identify as a molecular state, prior to the mixing induced by isospin breaking. Consequently, $E=-B$ is a zero of the $\Pi_{22}^{-1}$ matrix element, meaning it is a solution to the equation
\begin{equation}
    B+m_T-\frac{m}{2\pi}g_T^2(B)\left(\sqrt{2mB}+\sqrt{2m(B+\Delta)}\right)=0\,.
\end{equation}

In the absence of mixing ($g_{\text{mix}}=0$), the physical states would correspond exactly to the strong isospin eigenstates. The actual physical states, however, correspond to the poles of the full propagator
\begin{multline}
    \Pi(E)=\frac{1}{D(E)}\\\begin{pmatrix}
        -E+m_T-\dfrac{m}{2\pi}g_T^2\left(\kappa_0(E)+\kappa_\pm(E)\right)+i\dfrac{\Gamma_T}{2} & -g_{\text{mix}}\\
        - g_{\text{mix}} & E-m_S-\dfrac{m}{2\pi}g_S^2\left(\kappa_0(E)+\kappa_\pm(E)\right)+i\dfrac{\Gamma_S}{2}
    \end{pmatrix}\,,
    \label{eq:prop}
\end{multline}
hence to the zero of the denominator $D(E)$, which is the determinant of $\Pi^{-1}(E)$
\begin{equation}
    D(E)=s_Ss_T-g_{\text{mix}}^2\,,
\end{equation}
where 
\begin{equation}
    s_S=E-m_S-\dfrac{m}{2\pi}g_S^2\left(\kappa_0(E)+\kappa_\pm(E)\right)+i\dfrac{\Gamma_S}{2}\,,
\end{equation}
\begin{equation}
    s_T=-E+m_T-\dfrac{m}{2\pi}g_T^2\left(\kappa_0(E)+\kappa_\pm(E)\right)+i\dfrac{\Gamma_T}{2}\,.
\end{equation}

\subsection{\texorpdfstring{Decay amplitude: $\mathcal{D}^{\text{decay}}$}{Decay amplitude: Ddecay}}
\label{sec:decay}

\subsubsection{\texorpdfstring{$DD^*$ decay}{DD decay}}
\label{sec:decayDD}
Regarding the decay amplitude, we distinguish between two scenarios, starting with the decay into $(D^{0}\bar{D}^{*0},\,D^{+}D^{*-})$, and then moving on to $J/\psi+\text{pions}$ in the next section. In this case, the decay matrix is simply given by (see Eqs.~\eqref{eq:lag_S}-\eqref{eq:lag_T}):
\begin{equation}
    \mathcal{D}^{\text{decay},{DD^*}}=\begin{pmatrix}
        g_S & g_T\\
        -g_S & g_T
    \end{pmatrix}
    \label{eq:matrix_gsgt}
\end{equation}
where rows $a=1$ and $a=2$ correspond to the neutral and charged channels, respectively. Using our factorization hypothesis in Eq.~\eqref{eq:A_tot}, the decay amplitude into the final state of charmed mesons reads
\begin{equation}
    \mathcal{A}_a^{DD^*}(E)=\sum_{i,\,j}\mathcal{D}^{\text{decay},{DD^*}}_{ai}\Pi(E)_{ij}\mathcal{C}_j^\text{prod}=c_S\sum_{i}\,\mathcal{D}^{\text{decay},DD^*}_{ai}\Pi(E)_{i1}\,.
\end{equation}
Explicitly
\begin{equation}
    \mathcal{A}^{DD^*}(E)=\frac{c_S}{D(E)}\begin{pmatrix}
        g_S\,s_T-g_{\text{mix}}\,g_T\\
        -g_S\,s_T-g_{\text{mix}}\,g_T
    \end{pmatrix}\,.
    \label{eq:A_DD}
\end{equation}
To linear order in $g_{\text{mix}}$, this expression reduces to the more familiar form
\begin{equation}
    \mathcal{A}^{DD^*}(E)\simeq c_S\begin{pmatrix}
        \dfrac{g_S}{s_S}-\dfrac{g_{\text{mix}}\,g_T}{s_Ss_T}\\[9pt]
        -\dfrac{g_S}{s_S}-\dfrac{g_{\text{mix}}\,g_T}{s_Ss_T}
    \end{pmatrix}\,.
\end{equation}

We now expand on what was anticipated in the \hyperref[sec:intro]{Introduction} regarding the findings in \cite{LHCb:2024vfz}, where a contribution from a $1^{++}$ state was observed near the $D^{\pm}D^{*\mp}$ threshold. If this contribution is associated with the $X(3872)$, it yields $\text{Br}(B^+\to K^+ X)\text{Br}(X\to D^{+}D^{*-})=(1.48^{+0.41}_{-0.35})\times 10^{-4}$, compared to a neutral channel contribution of $\text{Br}(B^+\to K^+ X)\text{Br}(X\to D^{0}\bar{D}^{*0},\bar{D}^{0}D^{*0})= (0.80\pm0.23)\times 10^{-4}$ \cite{Belle:2008fma, ParticleDataGroup:2024cfk}. These values are difficult to accommodate within a single-resonance scenario, as the charged channel is phase-space suppressed due to its threshold being $\Delta\simeq 8\,\text{MeV}$ above the neutral one, unless one assumes another strong source of isospin violation causing the $X(3872)$ to couple much more strongly to the charged pair than to the neutral one. 

Introducing an isotriplet partner provides a natural explanation for this discrepancy by interfering destructively in the neutral channel and constructively in the charged one. Such a mechanism is feasible because the singlet and triplet states couple with opposite relative signs in the two channels, as evident from Eq.~\eqref{eq:A_DD}. It is also worth noting that this interference is due to the mixing and vanishes in the isospin limit (i.e., $g_{\text{mix}}=0$). 

\subsubsection{\texorpdfstring{$J/\psi+\text{pions}$ decay}{J/psi decay}}
\label{sec:decayJpsi}

We now turn to the $(J/\psi\,3\pi,J/\psi\, 2\pi)$ final states. To properly account for the strong final-state rescattering of the pions, alongside the isospin-violating contributions induced by $\rho-\omega$ mixing, we model the decay amplitude relying on the $\mathcal{P}$-vector approach~\cite{ParticleDataGroup:2024cfk,LHCb:2022jez,Chung:1995, Aitchison:1972ay}. Such mixing effects have been experimentally probed for the $X(3872)$ by analyzing the dipion invariant mass spectrum~\cite{LHCb:2022jez}. 

We adopt a basis where the $\rho$ and $\omega$ vector mesons are states of definite isospin, mixing exclusively via their kinetic terms. Consequently, the inverse propagator matrix for the $\rho-\omega$ system takes the form~\cite{OConnell:1995nse}
\begin{equation}
    \Pi^{\rho\omega}_{ab}(m_{\text{lh}}) = \begin{pmatrix}
        \dfrac{1}{m_{\text{lh}}^2-m_\omega^2} & \dfrac{g_{\rho\omega}}{(m_{\text{lh}}^2-m_\rho^2)(m_{\text{lh}}^2-m_\omega^2)}\\
        \dfrac{g_{\rho\omega}}{(m_{\text{lh}}^2-m_\rho^2)(m_{\text{lh}}^2-m_\omega^2)} & \dfrac{1}{m_{\text{lh}}^2-m_\rho^2}
    \end{pmatrix}
\end{equation}
with $m_{\text{lh}}$ representing the invariant mass of the \textit{light hadron} system (either $\pi^+\pi^-$ or $\pi^+\pi^-\pi^0$).

The construction of the $\mathcal{P}$-vector involves two primary components: an annihilation matrix describing the $X_i \to J/\psi V$ transitions ($V = \rho, \omega$), and a decay matrix governing $V \to 2\pi/3\pi$. Under the assumption of exact isospin conservation in the $X_i\to J/\psi V$ vertices, the $X_S$ state couples exclusively to $J/\psi \omega$, whereas the $X_T^0$ state decays solely into $J/\psi \rho^0$. As a result, the annihilation matrix is diagonal in our chosen basis
\begin{equation}
    \mathcal{C}^{\text{ann}} = \begin{pmatrix}
        c_\omega & 0 \\
        0 & c_\rho
    \end{pmatrix}\,.
\end{equation}
Similarly to the production amplitudes, we treat these coefficients as approximately constant, neglecting their energy dependence. The matrix encoding the decays into the pionic final states is also diagonal
\begin{equation}
    \mathcal{C}^{\text{decay}} = \begin{pmatrix}
        g_{\omega\to 3\pi} & 0 \\
        0 & g_{\rho\to 2\pi}
    \end{pmatrix}\,.
\end{equation}
We can thus assemble the $\mathcal{P}$-vector as follows
\begin{equation}
    \mathcal{P}_{ai}(m_{\text{lh}}) =\sum_{b,\,c} \mathcal{C}^{\text{decay}}_{ab} \Pi^{\rho\omega}_{bc}(m_{\text{lh}}) \mathcal{C}^{\text{ann}}_{ci}\,.
\end{equation}

To fully describe the system, we must also specify the $\mathcal{K}$-matrix, which captures the rescattering of the pions mediated by the intermediate $\rho$ and $\omega$ resonances. It is given by the expression
\begin{equation}
    \mathcal{K}_{ab}(m_{\text{lh}}) =\sum_{c,\,d} \mathcal{C}^{\text{decay}}_{ac} \Pi^{\rho\omega}_{cd}(m_{\text{lh}}) \mathcal{C}^{\text{decay}}_{db}\,.
\end{equation}
Structurally, this factorization is completely analogous to the $B$-decay amplitude in Eq.~\eqref{eq:A_tot}.

The complete decay amplitude is ultimately formulated as~\cite{Aitchison:1972ay,ParticleDataGroup:2024cfk}
\begin{equation}
    \mathcal{D}^{\text{decay},n\pi}_{ai}(m_{\text{lh}}) =\sum_{b,\,i}\, [1-i \mathcal{K} (m_{\text{lh}})\varrho(m_{\text{lh}}^2)]^{-1}_{ab} \mathcal{P}_{bi}(m_{\text{lh}}) \,,
    \label{eq:D-decay}
\end{equation}
in which $\varrho = \mathrm{diag}(\varrho_{3\pi}, \varrho_{2\pi})$ denotes the diagonal phase-space matrix, explicitly detailed in Appendix~\ref{app:phase_space}.

By performing a first-order expansion in the mixing parameter $\epsilon=g_{\rho\omega}/(m_\omega^2-m_\rho^2)$~\cite{OConnell:1995nse}, the expression simplifies to
\begin{equation}\label{eq:Ddec}
    \mathcal{D}^{\text{decay,}n\pi}= 
    \left[ \begin{pmatrix}
        c_\omega\dfrac{g_{\omega\to 3\pi}}{s_\omega} & 0 \\
        0 &  c_\rho\dfrac{g_{\rho\to2\pi}}{s_\rho}
    \end{pmatrix} + g_{\rho\omega}
    \begin{pmatrix}
        0 & c_\rho\dfrac{g_{\omega\to 3\pi}}{s_\rho s_\omega} \\
        c_\omega\dfrac{g_{\rho\to 2\pi}}{s_\rho s_\omega} & 0
    \end{pmatrix}\right]+\mathcal{O}(g_{\text{mix}}^2)
    \,,
\end{equation}
Here, we have defined the full inverse propagators
\begin{align}
    s_\rho =  m_\rho^2-m^2_{\text{lh}} -i g_{\rho \to 2\pi}^2\,\varrho_{2\pi}(m_{\text{lh}}^2) , \\
    s_\omega =  m_\omega^2 - m^2_{\text{lh}} -i g^2_{\omega \to3 \pi}\,\varrho_{3\pi}(m_{\text{lh}}^2).
\end{align}
The effective couplings $g_{\rho\to2\pi}$ and $g_{\omega\to3\pi}$ are constrained by the physical decay widths via the relations
\begin{equation}
    g^2_{\rho\to2\pi}
    =\frac{m_\rho\,\Gamma_\rho}{\varrho(m_\rho^2)}\simeq1.20,\qquad
    g^2_{\omega\to3\pi}=\frac{m_\omega\,\Gamma_{\omega}\,\text{Br}(\omega\to3\pi)}{\varrho(m_\omega^2)}\simeq6.57\,\,
\end{equation}
using the current experimental values~\cite{ParticleDataGroup:2024cfk}
\begin{equation}
    m_\rho= 775.26\,\text{MeV},\quad\Gamma_\rho = 147.4\text{MeV} 
\end{equation}
alongside
\begin{equation}
    m_\omega= 782.66\,\text{MeV},\quad\Gamma_\omega = 8.68\text{MeV},\quad \text{Br}(\omega\to3\pi)=89.2\%\,. 
\end{equation}
For the $\rho$--$\omega$ mixing strength, we adopt the value
\begin{equation}
    g_{\rho\omega}=-4520\,\text{MeV}^2
\end{equation}
following Ref.~\cite{OConnell:1995nse}.

Consequently, the full amplitude describing the process $B^+\to K^+(X_i\to J/\psi+\text{pions})$ reads
\begin{equation}
    \mathcal{A}_a^{J/\psi n\pi}(E,m_{\text{lh}})=\sum_{i,\,j}\mathcal{D}^{\text{decay},n\pi}_{ai}(m_{\text{lh}})\Pi_{ij}(E)\mathcal{C}_{j}^{\text{prod}}=c_S\sum_{i}\,\mathcal{D}^{\text{decay},n\pi}_{ai}(m_{\text{lh}})\Pi_{i1}(E)
    \label{eq:A_pipi}
\end{equation}
where the index $a=1,2$ refers to the $3\pi$ and $2\pi$ channels, respectively.

For example, building on the formalism developed above, the differential lineshape for the $B^+\to K^+(X\to J/\psi \pi^+\pi^-)$ process reads
\begin{multline}
    \frac{dR\left(B^+\to K^+ J/\psi \pi^+\pi^-\right)}{dm_{J/\psi\pi^+\pi^-}dm_{\pi^+\pi^-}} \propto\,
    p(m_{J/\psi\pi^+\pi^-}^2,m_{J/\psi}^2,m_{\pi^+\pi^-}^2)\,
    p_\pi(m_{\pi^+\pi^-}^2)\\B_1\left(p_\pi(m^2_{\pi^+\pi^-})\right)
    \left|\mathcal{A}_{2}^{J/\psi n\pi}(E, m_{\pi^+\pi^-})\right|^2,
    \label{eq:diff_lineshape}
\end{multline}
where $m_{J/\psi\pi^+\pi^-}=E+m_{D^0}+m_{D^{*0}}$ and $p$ is the standard K\"allén function
\begin{equation}
    p(m_1^2,m_2^2,m_3^2)=\frac{\sqrt{m_1^4+m_2^4+m_3^4-2m_1^2m_2^2-2m_1^2m_3^2-2m_2^2m_3^2}}{2m_1}\,.
    \label{eq:kallen}
\end{equation}
The remaining phase-space components, $p_\pi$ and $B_1$, are defined in Appendix~\ref{app:phase_space}. The overall normalization factor has been dropped since it consists of energy independent couplings that do not alter the differential shape. For experimental purposes, this global scale is irrelevant as it cancels out in ratios of decay rates and is treated as a free overall yield parameter in lineshape fits. In Appendix~\ref{app:isospin}, we show how this framework can be employed for the analysis of strong isospin breaking, along the lines of the study conducted in \cite{LHCb:2022jez}.

\section{Qualitative lineshape analysis}
It is not a priori evident that the proposed model possesses the required flexibility to accommodate the diverse properties of the $X(3872)$. In this section, we aim to demonstrate that our framework is not immediately inconsistent with the known experimental signatures, and that it can qualitatively reproduce the features of the $X(3872)$ lineshape. Performing a direct fit to experimental data is currently beyond our reach. Furthermore, we do not consider it appropriate to simply use the branching ratios reported by the PDG \cite{ParticleDataGroup:2024cfk}, as these values were extracted from data under the assumption of a single-resonance model and a specific lineshape parameterization. 

Performing a comprehensive scan over the entire parameter space is beyond the scope of this work. Therefore, we fix the parameters to the representative values listed in Table~\ref{tab:par}, and we vary only $g_{\text{mix}}$ to illustrate its effect on the lineshape.

\begin{table}[h]
    \centering
    \begin{tabular}{|c|c|c|c|c|c|}
    \hline
         $m_S\,(\text{MeV})$ & $m_T\,(\text{MeV})$ & $\Gamma_S\,(\text{keV})$ & $\Gamma_T\,(\text{keV})$ & $|g_S|\,(\text{MeV}^{-1/2})$ & $|g_T|\,(\text{MeV}^{-1/2})$   \\
         \hline\hline
         $-1.59$ & 3.07 & $20$ & $40$ & $1\times10^{-2}$ & $1.2\times 10^{-2}$ \\
         \hline
    \end{tabular}
    \caption{Parameter values used in this section.}
    \label{tab:par}
\end{table}

\begin{figure}[t]
    \centering
    \begin{minipage}{0.49\textwidth}
        \centering
        \includegraphics[width=\linewidth]{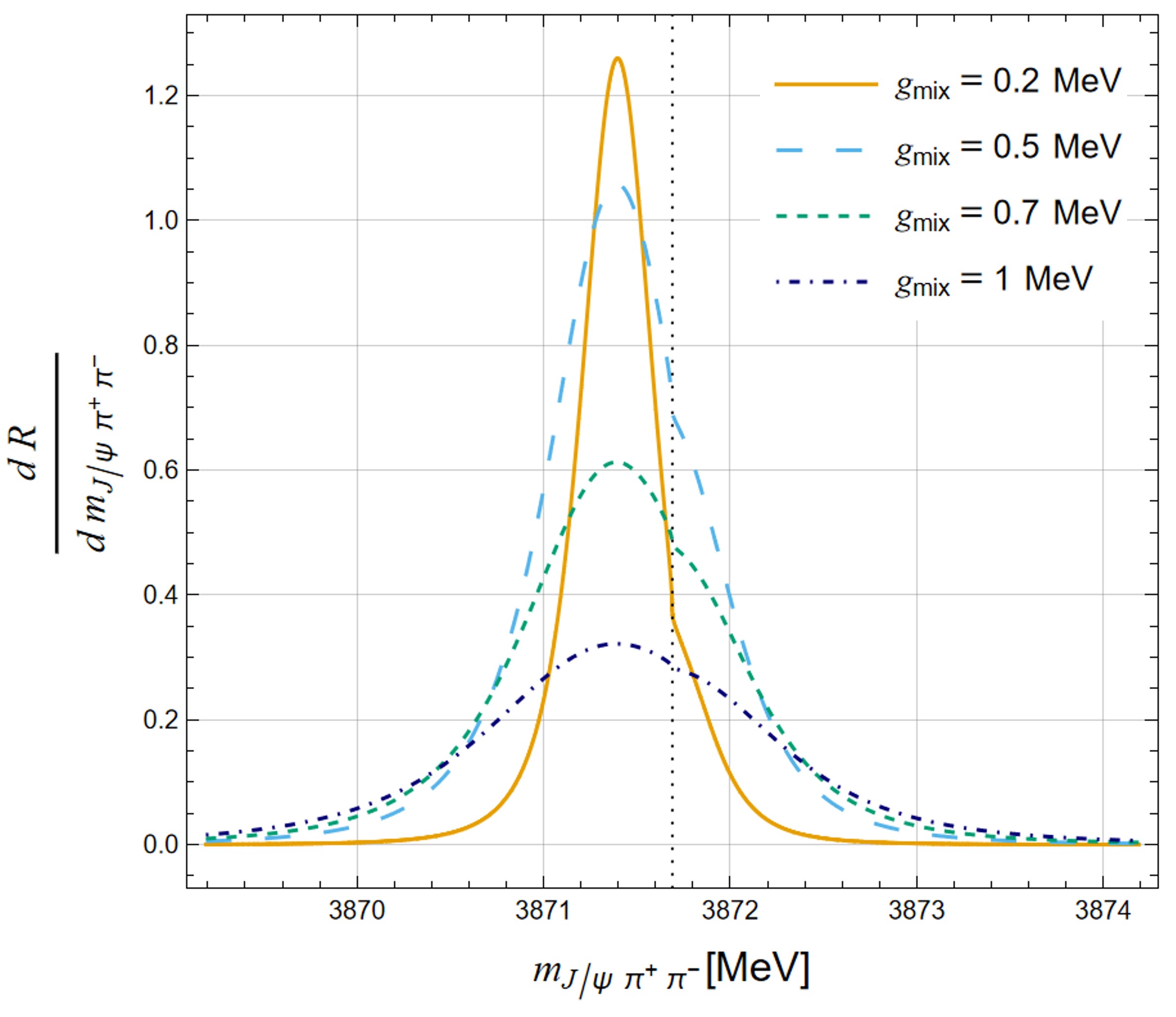}
        \caption{$J/\psi\pi^+\pi^-$ channel lineshape for different values of $g_{\text{mix}}$, in arbitrary units. The curve for $g_{\text{mix}}=0.2\,\text{MeV}$ is rescaled for better visualization. The vertical black dotted line indicates the $D^0D^{*0}$ threshold.}
        \label{fig:2pions}
    \end{minipage}\hfill
    \begin{minipage}{0.49\textwidth}
        \centering
        \includegraphics[width=\linewidth]{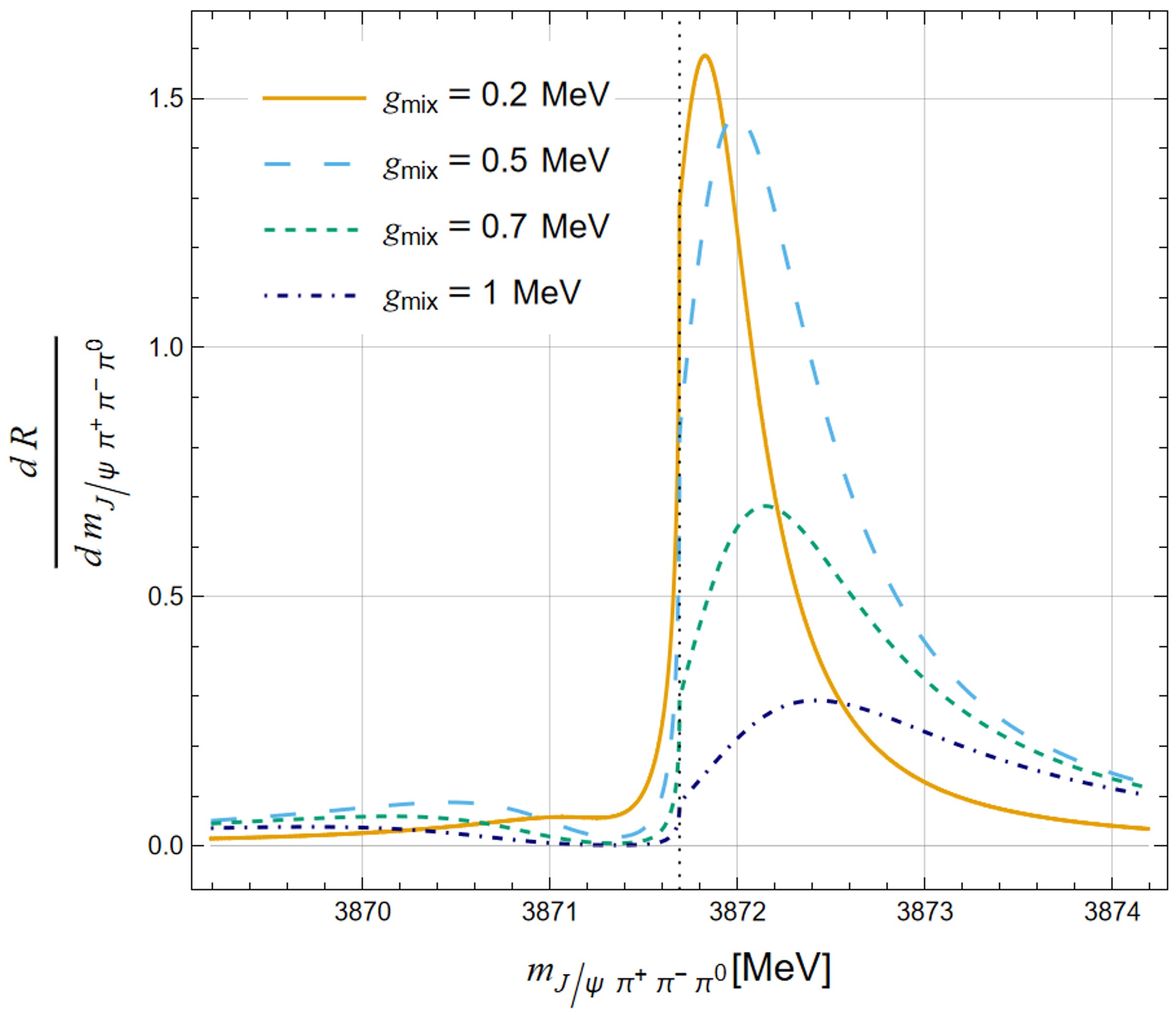}
        \caption{$J/\psi\pi^+\pi^-\pi^0$ channel lineshape for different values of $g_{\text{mix}}$, in arbitrary units. The curve for $g_{\text{mix}}=0.2\,\text{MeV}$ is rescaled for better visualization. The vertical black dotted line indicates the $D^0D^{*0}$ threshold.}
        \label{fig:3pions}
    \end{minipage}
\end{figure}

Let us consider, for instance, the process $B^+\to K^+ J/\psi+\text{pions}$, which is arguably the most illustrative for studying the $X(3872)$, as it coincides with the discovery channel of the resonance \cite{Belle:2003nnu} and is the focus of recent experimental analyses \cite{LHCb:2020xds,LHCb:2022jez}. For simplicity, we consider only the dominant $\rho-\omega$ mixing term
\begin{equation}
    D^{\text{decay},n\pi}\simeq \begin{pmatrix}
        c_\omega\dfrac{g_{\omega\to 3\pi}}{s_\omega} & 0 \\
        0 &  c_\rho\dfrac{g_{\rho\to2\pi}}{s_\rho}
    \end{pmatrix}\,.
\end{equation}
Hence, the decay amplitude takes the form
\begin{equation}
    \mathcal{A}^{J/\psi n\pi}(E,m_{\text{lh}})\simeq\frac{c_S}{D(E)}\begin{pmatrix}
        c_\omega \dfrac{g_{\omega\to3\pi}}{s_\omega}\left(-E+m_T-\dfrac{m}{2\pi}g_T^2\left(\kappa_0(E)+\kappa_\pm(E)\right)+i\dfrac{\Gamma_T}{2}\right)\\
        c_\rho \dfrac{g_{\rho\to2\pi}}{s_\rho}g_{\text{mix}}
    \end{pmatrix}\,.
\end{equation}
{To obtain the lineshape, we must multiply the squared decay amplitude by the two-body phase-space factor $p(m_{J/\psi+\text{pions}}^2, m_{J/\psi}^2, m_{\text{lh}}^2)$ --- the momentum of the $J/\psi$ in the rest frame of the intermediate state --- and by the phase-space factor for the vector meson decay into pions (i.e. $\varrho_{n\pi}(m_{\text{lh}}^2)$, explicitly written in Appendix~\ref{app:phase_space})}. Subsequently, we integrate over the invariant mass $m_{\text{lh}}$ to retain the dependence solely on $E$ (recall that $E$ is defined as $E = m_{\text{f.s.}} - m_{D^0} - m_{D^{*0}}$). This calculation is significantly simplified if we focus on the $m_{J/\psi+\text{pions}}$ region around the $D^{0}D^{*0}$ threshold. In this regime, the integral over the pions' invariant mass yields two distinct multiplicative constants that are independent of $E$. This is because the only term depending simultaneously on $E$ and $m_{\text{lh}}$ is the phase-space factor $p(m_{J/\psi+\text{pions}}^2,m_{J/\psi}^2,m_{\text{lh}}^2)$ (see, for example, Eq.~\eqref{eq:diff_lineshape}), which can be approximated as $p(m_{D^0+D^{*0}}^2,m_{J/\psi}^2,m_{\text{lh}}^2)$. Consequently, the lineshape $dR/dm_{J/\psi+\text{pions}}$ can be written as
\begin{multline}
    \frac{dR(B^+\to K^+J/\psi+\text{pions)}}{dm_{J/\psi+\text{pions}}}\propto\\\frac{1}{|D(E)|^2}\begin{pmatrix}
        \left|-E+m_T-\dfrac{m}{2\pi}g_T^2\left(\kappa_0(E)+\kappa_\pm(E)\right)+i\dfrac{\Gamma_T}{2}\right|^2\\
        |g_{\text{mix}}|^2
    \end{pmatrix}=\begin{pmatrix}
        |\Pi_{11}|^2\\
        |\Pi_{21}|^2
    \end{pmatrix}\,,
    \label{eq:amp_sim}
\end{multline}
where the first element of the column vector refers to the final state with three pions and the second to the one with two pions. In Figs.~\ref{fig:2pions} and \ref{fig:3pions}, we show the behavior of the lineshapes for both channels near the $D^0D^{*0}$ threshold as $g_{\text{mix}}$ varies. For the chosen parameter values, we obtain one complex pole of the propagator~\eqref{eq:prop} on the physical Riemann sheet $(+,+)$ below threshold, and another one slightly above threshold on the unphysical sheet $(-,+)$.\footnote{In our two-channel system, the Riemann sheets are conventionally denoted by $(s_1, s_2)$, where $s_1 = \text{sgn}\big(\Im\big(\sqrt{E}\big)\big)$ and $s_2 = \text{sgn}\big(\Im\big(\sqrt{E-\Delta}\big)\big)$ correspond to the neutral and charged channels, respectively. Thus, the $(+,+)$ configuration identifies the physical sheet, while $(-,+)$ indicates the unphysical sheet with respect to the $D^0\bar{D}^{*0}$ threshold \cite{ParticleDataGroup:2024cfk}.} As the value of $g_{\text{mix}}$ increases, the poles move deeper into the complex plane, leading to a broadening of the peaks, as seen in Fig.~\ref{fig:2pions}.

We wish to draw particular attention, however, to the peculiar behavior of the three-pion channel shown in Fig.~\ref{fig:3pions}. This arises because, in the first component of Eq.~\eqref{eq:amp_sim}, the numerator also possesses a zero on the physical sheet (the complex pole of the $X_T^0$ before the mixing) that lies closer to the real negative semiaxis than the pole of the denominator. This results in a nearly complete cancellation of the below-threshold peak and the emergence of a peak above threshold.\footnote{One might wonder whether this accidental cancellation is spoiled when the full $\rho-\omega$ mixing contribution is included. However, the effect of the additional terms is heavily suppressed upon phase-space integration due to the presence of the product $s_\rho s_\omega$, unless $c_\rho/c_\omega\sim\mathcal{O}(10^3)$.} As $g_{\text{mix}}$ increases, the zero in the numerator remains fixed while the zero of the denominator shifts in the complex plane. Consequently, the dip near the threshold becomes more pronounced, as can be observed in Fig.~\ref{fig:3pions}. Conversely, in the limit of vanishing $g_{\text{mix}}$, the numerator zero and the pole exactly cancel out, as $D(E)\to s_Ss_T$, and the lineshape no longer exhibits the below-threshold zero. While challenging to experimentally resolve, if observed, this feature would provide strong evidence in favor of our model. Such behavior cannot be reproduced in a single-resonance scenario, since this effect intrinsically relies on the non-diagonal structure of the propagator matrix $\Pi(E)$ \eqref{eq:prop}, which in the single-resonance case trivially reduces to $1/s_S$.

\begin{figure}[t]
    \centering
    \begin{minipage}{0.48\textwidth}
        \centering
        \includegraphics[width=\linewidth]{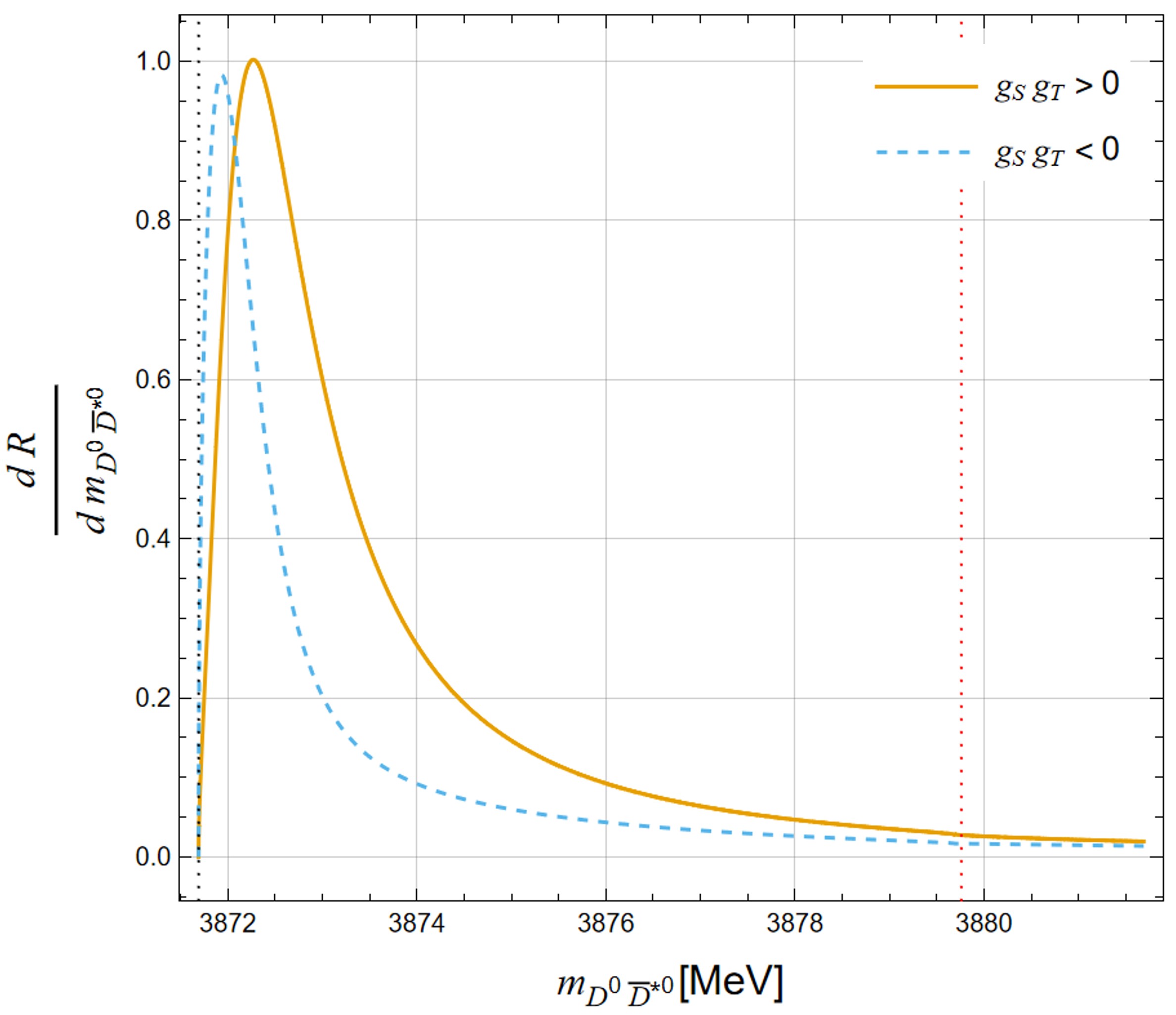}
        \caption{$D^0\bar{D}^{*0}$ channel lineshape for $g_{\text{mix}}=0.7\,\text{MeV}$, in arbitrary units. The vertical black dotted line indicates the $D^0\bar{D}^{*0}$ threshold, while the red one marks the $D^+D^{*-}$ threshold. The lineshapes are rescaled by a common factor.}
        \label{fig:d0d0}
    \end{minipage}\hfill
    \begin{minipage}{0.49\textwidth}
        \centering
        \includegraphics[width=\linewidth]{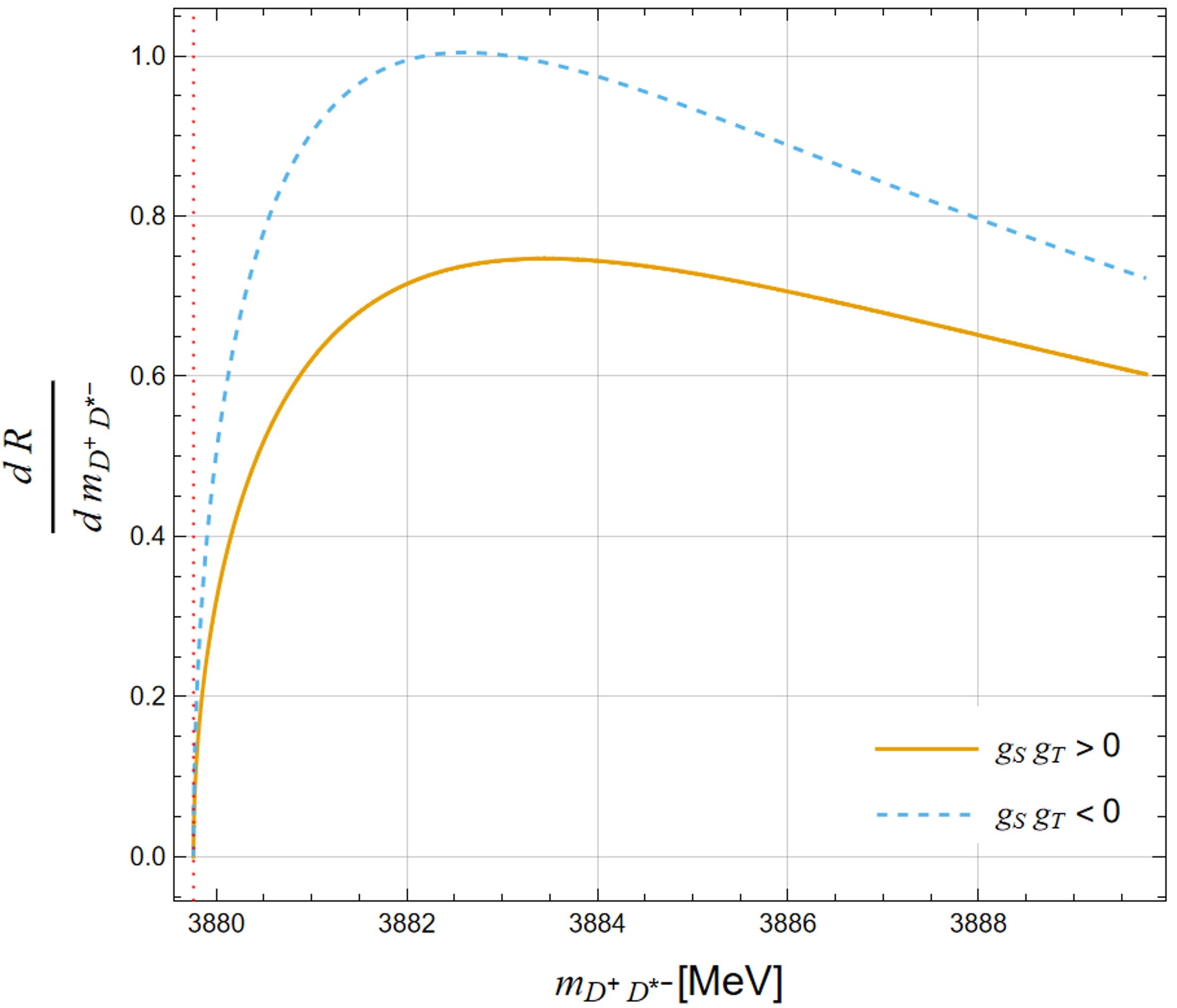}
        \caption{$D^+D^{*-}$ channel lineshape for $g_{\text{mix}}=0.7\,\text{MeV}$, in arbitrary units. The vertical black dotted line indicates the $D^0\bar{D}^{*0}$ threshold, while the red one marks the $D^+D^{*-}$ threshold. The lineshapes are rescaled by a common factor.}
        \label{fig:dpdm}
    \end{minipage}
\end{figure}

To fully appreciate the role of the relative sign between the couplings $g_i$, it is necessary to investigate the processes involving $DD^*$ mesons. We consider the lineshape of the $B^+\to K^+ DD^{*}$ decay, expressed in our model as
\begin{equation}
    \frac{dR\left(B^+\to K^+ DD^{*}\right)}{dm_{DD^*}}\propto p(m_{DD^{*}}^2,m^2_{D},m^2_{D^{*}})
    |\mathcal{A}_i^{DD^*}(m_{DD^{*}})|^2\,,
\end{equation}
where $i=1,2$ denotes the neutral or charged channel, and the final-state phase space has been properly included. In Fig.~\ref{fig:d0d0}, we display the neutral case for $g_{\text{mix}}=0.7\,\text{MeV}$ under the assumption of both equal and opposite signs for the $g_i$ couplings in Table \ref{tab:par}. The effect of the interference is reflected not only in the relative magnitude but also in the difference in the overall profiles; this could potentially allow for an experimental extraction of the relative sign between the couplings. Furthermore, in Fig.~\ref{fig:dpdm}, we show the outcome for the charged mesons around the $D^{+}D^{*-}$ threshold. Notably, flipping the relative sign leads to an enhancement of the charged lineshape, possibly achieving the behavior discussed in Sec. \ref{sec:decayDD}.

\section{Conclusions and future prospects}

We have presented a model in which  the experimentally observed $X(3872)$ is a mix of an isosinglet compact tetraquark state $X_S$ and a neutral isotriplet molecular partner $X_T^0$. It is therefore interpreted as just one of the two physical states resulting from the mixing of the $X_S$ and $X_T$ basis states. This mixing of states is induced by isospin breaking. A salient feature of this formalism is its ability to simultaneously describe the large observed isospin breaking across multiple decay channels within a unified framework. This provides a robust method for cross-checking the validity of the underlying dynamics, rather than relying on different, channel-specific models for each process. Furthermore, this framework is not in evident tension with existing experimental data; as discussed in the previous section, by employing representative parameters — consistent with \cite{Carducci:2025jed} — we have shown that it is possible to qualitatively reproduce the observed properties of the $X(3872)$. Indeed, it is important to stress that it is presently inconsistent to directly adopt the parameter values obtained from current experimental analyses \cite{LHCb:2020xds,LHCb:2024vfz,LHCb:2022jez}, as these values are extracted from data relying on a different lineshape parameterization that assumes a single-resonance scenario.

Looking ahead, the proposed formalism is highly versatile and opens up several future prospects. With minor modifications — specifically by explicitly incorporating the decay of the $D^{*0}$ — the model can naturally accommodate additional final states, such as $D^0\bar{D}^0\gamma$ and $D^0\bar{D}^0\pi^0$. Furthermore, it provides a solid foundation for performing a dedicated theoretical study of isospin violation in $J/\psi\pi^+\pi^-$ decays, offering a valuable counterpart to recent experimental investigations like the one reported in \cite{LHCb:2022jez}.

We strongly encourage a comprehensive, simultaneous experimental fit of the relevant decay channels utilizing the lineshape derived in this work. We believe that such a global analysis is the crucial next step toward obtaining a complete and coherent theoretical and experimental picture of the $X(3872)$.

\appendix
\section{Phase space decay factors}
\label{app:phase_space}

Since the vector mesons decay into the $\pi^+\pi^-$ system via a $P$-wave transition, it is necessary to incorporate the Blatt--Weisskopf centrifugal barrier factors~\cite{osti_5992854}. In alignment with the approach of Ref.~\cite{LHCb:2022jez}, we embed the explicit $|\bm{p}|^l$ momentum scaling corresponding to the $l=1$ partial wave directly into this factor
\begin{equation}
    B_1(|\bm{p}|)=\frac{|\bm{p}|^2}{1+(R|\bm{p}|)^2}\,,
\end{equation}
Here, the parameter $R$ characterizes the effective hadronic radius~\cite{LHCb:2022jez,PhysRevD.83.074004}.

Using this definition, the diagonal elements of the phase-space matrix introduced in Eq.~\eqref{eq:D-decay} evaluate to
\begin{equation}
    \varrho_{2\pi}(s)=\frac{2p_\pi(s)}{\sqrt{s}}\,B_1\big(p_\pi(s)\big)\,,
\end{equation}
with
\begin{equation}
    p_\pi(s)=p(s,m_{\pi^+}^2,m_{\pi^-}^2)=\frac{1}{2}\sqrt{s-(2m_{\pi^+})^2}
\end{equation}
representing the relative momentum of the outgoing pions evaluated in the $\pi^+\pi^-$ rest frame. 

Regarding the three-pion final state, the $\omega\to3\pi$ decay is governed primarily by an intermediate $\rho\pi$ state in a $P$-wave configuration. Integrating over this intermediate state yields~\cite{LHCb:2022jez}:
\begin{equation}
    \varrho_{3\pi}(s)
    =\int_{(2m_{\pi^+})^2}^{(\sqrt{s}-m_{\pi^+})^2}
    \frac{\varrho_{2\pi}(\sigma)}
    {(m_\rho^2-\sigma)^2+(m_\rho\Gamma_\rho)^2}
    \frac{2p(s,\sigma,m_{\pi^0}^2)}{\sqrt{s}}
    B_1\big(p(s,\sigma,m_{\pi^0}^2)\big)\,d\sigma\,.
\end{equation}
\section{Strong isospin violation}
\label{app:isospin}
In this appendix, we demonstrate how to implement the analysis of strong isospin violation by studying the $B^+\to K^+ J/\psi \pi^+\pi^-$ process. We adapt the methodology employed in \cite{LHCb:2022jez} to facilitate its integration into future experimental analyses. 

We recall that the lineshape for the $B^+\to K^+ J/\psi \pi^+\pi^-$ process is given by
\begin{multline}
    \frac{dR\left(B^+\to K^+ J/\psi \pi^+\pi^-\right)}{dm_{J/\psi\pi^+\pi^-}dm_{\pi^+\pi^-}} =\mathcal{N}\,
    p(m_{J/\psi\pi^+\pi^-}^2,m_{J/\psi}^2,m_{\pi^+\pi^-}^2)\,
    p_\pi(m_{\pi^+\pi^-}^2)\\B_1\left(p_\pi(m^2_{\pi^+\pi^-})\right)
    \left|\mathcal{A}_{2}^{J/\psi n\pi}(E, m_{\pi^+\pi^-})\right|^2
\end{multline}
where $\mathcal{N}$ is a nuisance parameter and $E=m_{J/\psi\pi^+\pi^-}-m_{D^0}-m_{D^{*0}}$. By integrating over the $m_{J/\psi\pi^+\pi^-}$ invariant mass around the $X(3872)$ pole, we obtain the $m_{\pi^+\pi^-}$ mass distribution. This distribution replaces the $\mathcal{P}(m_{\pi^+\pi^-})$ probability density function used to fit the data in \cite{LHCb:2022jez}.

At linear order in $g_{\rho\omega}$, the decay amplitude can be written as (see Eq.~\eqref{eq:A_pipi})
\begin{equation}
    \mathcal{A}_2^{J/\psi n\pi}(E,m_{\pi^+\pi^-})=\frac{c_S\,g_{\rho\to2\pi}}{D(E)}\left(c_\rho \frac{g_{\text{mix}}}{s_\rho}-c_\omega g_{\rho\omega}\frac{s_T}{s_\rho s_\omega}\right)\,.
\end{equation}
To estimate the relative contribution between the $\omega$ and $\rho$ mesons, we define the following quantities
\begin{equation}
    I_\rho=\int dm_{\pi^+\pi^-}\int dm_{J/\psi\pi^+\pi^-} \frac{dR(B^+\to K^+ J/\psi\pi^+\pi^-)}{dm_{J/\psi\pi^+\pi^-} dm_{\pi^+\pi^-}}\Bigg|_{c_{\omega}=0}
\end{equation}
and
\begin{equation}
    I_\omega=\int dm_{\pi^+\pi^-}\int dm_{J/\psi\pi^+\pi^-} \frac{dR(B^+\to K^+ J/\psi\pi^+\pi^-)}{dm_{J/\psi\pi^+\pi^-} dm_{\pi^+\pi^-}}\Bigg|_{c_{\rho}=0}\,.
\end{equation}
Isospin violation is related to the ratio
\begin{equation}
    \mathcal{R}_{\omega/\rho}=\frac{I_\omega}{I_\rho}\,,
\end{equation}
therefore, we can factor out and cancel all the overall constant terms between the numerator and the denominator in the integrals, yielding
\begin{multline}
    \mathcal{R}_{\omega/\rho}=\left|\frac{c_\omega}{c_\rho}\right|^2\\\dfrac{\displaystyle\int d m_{\pi^+\pi^-}\int dm_{J/\psi \pi^+\pi^- }p(m_{J/\psi\pi^+\pi^-}^2,m_{J/\psi}^2,m_{\pi^+\pi^-}^2)\,
    p_\pi(m_{\pi^+\pi^-}^2)B_1\left(p_\pi(m^2_{\pi^+\pi^-})\right)\left|\dfrac{g_{\rho\omega}s_T}{D(E)s_\rho s_\omega}\right|^2}{\displaystyle\int d m_{\pi^+\pi^-}\int dm_{J/\psi \pi^+\pi^- }p(m_{J/\psi\pi^+\pi^-}^2,m_{J/\psi}^2,m_{\pi^+\pi^-}^2)\,
    p_\pi(m_{\pi^+\pi^-}^2)B_1\left(p_\pi(m^2_{\pi^+\pi^-})\right)\left|\dfrac{g_{\text{mix}}}{D(E)s_\rho}\right|^2}\,.
\end{multline}

\acknowledgments

B.G.~thanks Antonio Polosa and Alessandro Melchiorri for stimulating lunch conversations. 
The work was developed and completed during B.G's  Visiting Professorship at Universit\`a di Roma, La Sapienza. The work of B.G. is
supported in part by the U.S. Department of Energy under grant number DE-SC0009919.

\newpage
\bibliographystyle{JHEP}
{\footnotesize
\bibliography{refs}}

@article{Esposito:2025hlp,
    author = "Esposito, Angelo and Glioti, Alfredo and Germani, Davide and Polosa, Antonio D.",
    title = "{A short review on the compositeness of the X(3872)}",
    eprint = "2502.02505",
    archivePrefix = "arXiv",
    primaryClass = "hep-ph",
    doi = "10.1007/s40766-025-00066-3",
    journal = "Riv. Nuovo Cim.",
    volume = "48",
    number = "2",
    pages = "95--155",
    year = "2025"
}

@article{LHCb:2022jez,
    author = "Aaij, Roel and others",
    collaboration = "LHCb",
    title = "{Observation of sizeable {\ensuremath{\omega}} contribution to $\chi_{c1}(3872)\to \pi^+\pi^- J/\psi$ decays}",
    eprint = "2204.12597",
    archivePrefix = "arXiv",
    primaryClass = "hep-ex",
    reportNumber = "LHCb-PAPER-2021-045, CERN-EP-2022-049",
    doi = "10.1103/PhysRevD.108.L011103",
    journal = "Phys. Rev. D",
    volume = "108",
    number = "1",
    pages = "L011103",
    year = "2023"
}

@article{Chung:1995,
    author = "Chung, S. U. and others",
    title = "{Partial wave analysis in K-matrix formalism}",
    doi = "10.1002/andp.19955070504",
    journal = "Annalen Phys.",
    volume = "4",
    number = "7",
    pages = "404--430",
    year = "1995"
}

@article{OConnell:1995nse,
    author = "O'Connell, Heath Bland and Pearce, B. C. and Thomas, Anthony William and Williams, Anthony Gordon",
    title = "{$\rho - \omega$ mixing, vector meson dominance and the pion form-factor}",
    eprint = "hep-ph/9501251",
    archivePrefix = "arXiv",
    reportNumber = "ADP-95-1-T-168",
    doi = "10.1016/S0146-6410(97)00044-6",
    journal = "Prog. Part. Nucl. Phys.",
    volume = "39",
    pages = "201--252",
    year = "1997"
}

@article{Aitchison:1972ay,
    author = "Aitchison, I. J. R.",
    title = "{K-MATRIX FORMALISM FOR OVERLAPPING RESONANCES}",
    reportNumber = "CLARENDON-1-72",
    doi = "10.1016/0375-9474(72)90305-3",
    journal = "Nucl. Phys. A",
    volume = "189",
    pages = "417--423",
    year = "1972"
}

@article{PhysRevD.83.074004,
  title = {Pion-pion scattering amplitude. IV. Improved analysis with once subtracted Roy-like equations up to 1100 MeV},
  author = {Garc\'{\i}a-Mart\'{\i}n, R. and Kami\ifmmode \acute{n}\else \'{n}\fi{}ski, R. and Pel\'aez, J. R. and Ruiz de Elvira, J. and Yndur\'ain, F. J.},
  journal = {Phys. Rev. D},
  volume = {83},
  issue = {7},
  pages = {074004},
  numpages = {34},
  year = {2011},
  month = {Apr},
  publisher = {American Physical Society},
  doi = {10.1103/PhysRevD.83.074004},
  url = {https://link.aps.org/doi/10.1103/PhysRevD.83.074004}
}

@book{osti_5992854,
  author       = {Blatt, J M and Weisskopf, V F},
  title        = {Theoretical nuclear physics},
  url          = {https://www.osti.gov/biblio/5992854},
  place        = {United States},
  publisher    = {Springer-Verlag New York Inc.,New York, NY},
  year         = {1978},
  month        = {12}
}

@article{Artoisenet:2010va,
    author = "Artoisenet, Pierre and Braaten, Eric and Kang, Daekyoung",
    title = "{Using Line Shapes to Discriminate between Binding Mechanisms for the X(3872)}",
    eprint = "1005.2167",
    archivePrefix = "arXiv",
    primaryClass = "hep-ph",
    doi = "10.1103/PhysRevD.82.014013",
    journal = "Phys. Rev. D",
    volume = "82",
    pages = "014013",
    year = "2010"
}

@article{LHCb:2024vfz,
    author = "Aaij, Roel and others",
    collaboration = "LHCb",
    title = "{Observation of New Charmonium or Charmoniumlike States in $B^+\to D^{*\pm}D^{\mp}K^+$ Decays}",
    eprint = "2406.03156",
    archivePrefix = "arXiv",
    primaryClass = "hep-ex",
    reportNumber = "LHCb-PAPER-2023-047, CERN-EP-2024-096",
    doi = "10.1103/PhysRevLett.133.131902",
    journal = "Phys. Rev. Lett.",
    volume = "133",
    number = "13",
    pages = "131902",
    year = "2024"
}

@unpublished{Carducci:2025jed,
    author = "Carducci, A. and Grinstein, B. and Germani, D. and Polosa, A. D.",
    title = "{The Compact X and Z and their Invisible Molecular Partners}",
    eprint = "2512.21538",
    archivePrefix = "arXiv",
    primaryClass = "hep-ph",
    month = "12",
    year = "2025"
}

@article{Kaplan:1999qa,
    author = "Kaplan, David B. and Steele, James V.",
    title = "{The Long and short of nuclear effective field theory expansions}",
    eprint = "nucl-th/9905027",
    archivePrefix = "arXiv",
    reportNumber = "CERN-TH-99-138, DOE-ER-40561-55-INT99",
    doi = "10.1103/PhysRevC.60.064002",
    journal = "Phys. Rev. C",
    volume = "60",
    pages = "064002",
    year = "1999"
}

@article{Kaplan:1996nv,
    author = "Kaplan, David B.",
    title = "{More effective field theory for nonrelativistic scattering}",
    eprint = "nucl-th/9610052",
    archivePrefix = "arXiv",
    reportNumber = "DOE-ER-40561-296, INT-96-00-153, UW-PT-96-31",
    doi = "10.1016/S0550-3213(97)00178-8",
    journal = "Nucl. Phys. B",
    volume = "494",
    pages = "471--484",
    year = "1997"
}

@article{Belle:2003nnu,
    author = "Choi, S. K. and others",
    collaboration = "Belle",
    title = "{Observation of a narrow charmonium-like state in exclusive $B^\pm \to K^\pm \pi^+ \pi^- J/\psi$ decays}",
    eprint = "hep-ex/0309032",
    archivePrefix = "arXiv",
    doi = "10.1103/PhysRevLett.91.262001",
    journal = "Phys. Rev. Lett.",
    volume = "91",
    pages = "262001",
    year = "2003"
}

@article{ParticleDataGroup:2024cfk,
    author = "Navas, S. and others",
    collaboration = "Particle Data Group",
    title = "{Review of particle physics}",
    doi = "10.1103/PhysRevD.110.030001",
    journal = "Phys. Rev. D",
    volume = "110",
    number = "3",
    pages = "030001",
    year = "2024",
    note= "The analytical structure of the amplitude can be found in Sec. 50.1.1. The $\mathcal{K}-$matrix and $\mathcal{P}-$vector formalism can be found in Sec. 50.3.2 \textit{$\mathcal{K}$-matrix approach}"
}

@article{Esposito:2015fsa,
    author = "Esposito, A. and Guerrieri, A. L. and Maiani, L. and Piccinini, F. and Pilloni, A. and Polosa, A. D. and Riquer, V.",
    title = "{Observation of light nuclei at ALICE and the $X(3872)$ conundrum}",
    eprint = "1508.00295",
    archivePrefix = "arXiv",
    primaryClass = "hep-ph",
    doi = "10.1103/PhysRevD.92.034028",
    journal = "Phys. Rev. D",
    volume = "92",
    number = "3",
    pages = "034028",
    year = "2015"
}

@article{CMS_pp_ptproduction,
    author = "Chatrchyan, Serguei and others",
    collaboration = "CMS",
    title = "{Measurement of the $X$(3872) Production Cross Section Via Decays to $J/\psi \pi^+ \pi^-$ in $pp$ collisions at $\sqrt{s}$ = 7 TeV}",
    eprint = "1302.3968",
    archivePrefix = "arXiv",
    primaryClass = "hep-ex",
    reportNumber = "CMS-BPH-11-011, CERN-PH-EP-2013-014",
    doi = "10.1007/JHEP04(2013)154",
    journal = "JHEP",
    volume = "04",
    pages = "154",
    year = "2013"
}

@article{LHCb_High_mult_pp_collision,
    author = "Aaij, Roel and others",
    collaboration = "LHCb",
    title = "{Observation of Multiplicity Dependent Prompt $\chi_{c1}(3872)$ and $\psi(2S)$ Production in $pp$ Collisions}",
    eprint = "2009.06619",
    archivePrefix = "arXiv",
    primaryClass = "hep-ex",
    reportNumber = "LHCb-PAPER-2020-023, CERN-EP-2020-161",
    doi = "10.1103/PhysRevLett.126.092001",
    journal = "Phys. Rev. Lett.",
    volume = "126",
    number = "9",
    pages = "092001",
    year = "2021"
}

@article{CMS_pbpb_ptproduction,
    author = "Sirunyan, Albert M. and others",
    collaboration = "CMS",
    title = "{Evidence for X(3872) in Pb-Pb Collisions and Studies of its Prompt Production at $\sqrt {s_{NN}}$=5.02\,\,TeV}",
    eprint = "2102.13048",
    archivePrefix = "arXiv",
    primaryClass = "hep-ex",
    reportNumber = "CMS-HIN-19-005, CERN-EP-2021-023",
    doi = "10.1103/PhysRevLett.128.032001",
    journal = "Phys. Rev. Lett.",
    volume = "128",
    number = "3",
    pages = "032001",
    year = "2022"
}

@article{LHCb:2020xds,
    author = "Aaij, R. and others",
    collaboration = "LHCb",
    title = "{Study of the lineshape of the $\chi_{c1}(3872)$ state}",
    eprint = "2005.13419",
    archivePrefix = "arXiv",
    primaryClass = "hep-ex",
    reportNumber = "CERN-EP-2020-086, LHCb-PAPER-2020-008",
    doi = "10.1103/PhysRevD.102.092005",
    journal = "Phys. Rev. D",
    volume = "102",
    number = "9",
    pages = "092005",
    year = "2020"
}

@article{landau,
    author = "Landau, Lev Davidovich",
    editor = "ter Haar, D.",
    title = "{Small Binding Energies in Quantum Field Theory}",
    doi = "10.1016/b978-0-08-010586-4.50104-8",
    journal = "J. Exp. Theor. Phys.",
    volume = "39",
    year = "1960"
}

@article{Bignamini:2009sk,
    author = "Bignamini, C. and Grinstein, B. and Piccinini, F. and Polosa, A. D. and Sabelli, C.",
    title = "{Is the X(3872) Production Cross Section at Tevatron Compatible with a Hadron Molecule Interpretation?}",
    eprint = "0906.0882",
    archivePrefix = "arXiv",
    primaryClass = "hep-ph",
    doi = "10.1103/PhysRevLett.103.162001",
    journal = "Phys. Rev. Lett.",
    volume = "103",
    pages = "162001",
    year = "2009"
}

@book{DWBA,
  title={Lectures on quantum mechanics},
  author={Weinberg, Steven},
  year={2015},
  publisher={Cambridge University Press}
}

@inproceedings{Belle:2005lfc,
    author = "Abe, Kazuo and others",
    collaboration = "Belle",
    title = "{Evidence for $X(3872)$ ---{\ensuremath{>}} $\gamma\,J / \psi$ and the sub-threshold decay $X(3872)$ ---{\ensuremath{>}} $\omega\,J / \psi$}",
    booktitle = "{22nd International Symposium on Lepton-Photon Interactions at High Energy (LP 2005)}",
    eprint = "hep-ex/0505037",
    archivePrefix = "arXiv",
    reportNumber = "BELLE-CONF-0540, LP-2005-175",
    month = "5",
    year = "2005"
}

@article{BESIII:2019esk,
    author = "Ablikim, M. and others",
    collaboration = "BESIII",
    title = "{Observation of the decay $X(3872) \to \pi^0 \chi_{c1}(1P)$}",
    eprint = "1901.03992",
    archivePrefix = "arXiv",
    primaryClass = "hep-ex",
    doi = "10.1103/PhysRevLett.122.202001",
    journal = "Phys. Rev. Lett.",
    volume = "122",
    number = "20",
    pages = "202001",
    year = "2019"
}

@article{BaBar:2010wfc,
    author = "del Amo Sanchez, P. and others",
    collaboration = "BaBar",
    title = "{Evidence for the decay $X(3872)$ ---{\ensuremath{>}} $J/\psi\,\omega$}",
    eprint = "1005.5190",
    archivePrefix = "arXiv",
    primaryClass = "hep-ex",
    reportNumber = "SLAC-PUB-14143, BABAR-PUB-10-009",
    doi = "10.1103/PhysRevD.82.011101",
    journal = "Phys. Rev. D",
    volume = "82",
    pages = "011101",
    year = "2010"
}

@article{Belle:2008fma,
    author = "Aushev, T. and others",
    collaboration = "Belle",
    title = "{Study of the $B\to X(3872)(D^{*0}\bar{D}^0) K$ decay}",
    eprint = "0810.0358",
    archivePrefix = "arXiv",
    primaryClass = "hep-ex",
    reportNumber = "BELLE-CONF-0832",
    doi = "10.1103/PhysRevD.81.031103",
    journal = "Phys. Rev. D",
    volume = "81",
    pages = "031103",
    year = "2010"
}

@article{Braaten:2006sy,
    author = "Braaten, Eric and Lu, Meng",
    title = "{Operator Product Expansion in the Production and Decay of the X(3872)}",
    eprint = "hep-ph/0606115",
    archivePrefix = "arXiv",
    doi = "10.1103/PhysRevD.74.054020",
    journal = "Phys. Rev. D",
    volume = "74",
    pages = "054020",
    year = "2006"
}

@article{Braaten:2005jj,
    author = "Braaten, Eric and Kusunoki, Masaoki",
    title = "{Factorization in the production and decay of the X(3872)}",
    eprint = "hep-ph/0506087",
    archivePrefix = "arXiv",
    doi = "10.1103/PhysRevD.72.014012",
    journal = "Phys. Rev. D",
    volume = "72",
    pages = "014012",
    year = "2005"
}

@unpublished{Brambilla:2026ujo,
    author = "Brambilla, Nora and Butenschoen, Mathias and Hibler, Simon and Mohapatra, Abhishek and Vairo, Antonio and Wang, Xiangpeng",
    title = "{Inclusive hadroproduction of $\chi_{c1}(3872)$, $X_b$ and pentaquarks}",
    eprint = "2602.14916",
    archivePrefix = "arXiv",
    primaryClass = "hep-ph",
    reportNumber = "TUM-EFT 203/26",
    month = "2",
    year = "2026"
}

@article{BaBar:2004cah,
    author = "Aubert, Bernard and others",
    collaboration = "BaBar",
    title = "{Search for a charged partner of the X(3872) in the $B$ meson decay $B \to X^- K$, $X^- \to J/\psi \pi^- \pi^0$}",
    eprint = "hep-ex/0412051",
    archivePrefix = "arXiv",
    reportNumber = "SLAC-PUB-10903, BABAR-PUB-04-043",
    doi = "10.1103/PhysRevD.71.031501",
    journal = "Phys. Rev. D",
    volume = "71",
    pages = "031501",
    year = "2005"
}

@unpublished{Carducci:2026fbz,
    author = "Carducci, A. and Cianti, G. and D'Annibali, P. and Germani, D. and Polosa, A. D.",
    title = "{A New Look at the X Compositeness from its Lineshape}",
    eprint = "2605.12274",
    archivePrefix = "arXiv",
    primaryClass = "hep-ph",
    month = "5",
    year = "2026"
}

@unpublished{Ji:2025hjw,
    author = "Ji, Teng and Dong, Xiang-Kun and Guo, Feng-Kun and Hanhart, Christoph and Mei{\ss}ner, Ulf-G.",
    title = "{Precise determination of the properties of $X(3872)$ and of its isovector partner $W_{c1}$}",
    eprint = "2502.04458",
    archivePrefix = "arXiv",
    primaryClass = "hep-ph",
    month = "2",
    year = "2025"
}

\end{document}